\documentclass[a4paper]{article}
\usepackage[a4paper, margin=2.5cm]{geometry}

\usepackage{graphicx}%
\usepackage{caption}
\usepackage{subcaption}%
\usepackage{multirow}%
\usepackage{amsmath,amssymb,amsfonts}%
\usepackage{amsthm}%
\usepackage{mathrsfs}%
\usepackage[title]{appendix}%
\usepackage{xcolor}%
\usepackage{textcomp}%
\usepackage{booktabs}%
\usepackage{algorithm}%
\usepackage{algorithmicx}%
\usepackage{algpseudocode}%
\usepackage{pdflscape} \pdfminorversion=7
\usepackage{url}

\usepackage[affil-it]{authblk}

\usepackage{natbib}

\usepackage{listings}%
\usepackage[scaled=0.9]{inconsolata}

\definecolor{codegreen}{rgb}{0,0.6,0}
\definecolor{codegray}{rgb}{0.5,0.5,0.5}
\definecolor{codeblue}{rgb}{0.0,0,0.82}
\definecolor{backcolour}{rgb}{0.95,0.95,0.92}

\lstdefinestyle{mystyle}{
    backgroundcolor=\color{backcolour},   
    commentstyle=\color{codegreen},
    numberstyle=\tiny\color{codegray},
    stringstyle=\color{codeblue},
    basicstyle=\ttfamily\small,
    breakatwhitespace=false,         
    breaklines=true,                 
    captionpos=b,                    
    keepspaces=true,                 
    numbers=left,                    
    numbersep=5pt,                  
    showspaces=false,                
    showstringspaces=false,
    showtabs=false,                  
    tabsize=2
}

\lstnewenvironment{rc}[1][]{\lstset{language=R}}{}

\lstdefinestyle{Routput}{
  basicstyle=\ttfamily\small,
  backgroundcolor=\color{backcolour},
  frame=single,
  rulecolor=\color{codegray},
  breaklines=true
}

\begin{document}

\title{\texttt{TwoTimeScales}: An R-package for Smoothing Hazards with Two Time Scales}

\author[1,3]{Angela Carollo}
\author[2]{Paul H.C. Eilers}
\author[3]{Hein Putter}
\author[1]{Jutta Gampe}

\affil[1]{Max Planck Institute for Demographic Research}
\affil[2]{Erasmus University Medical Center}
\affil[3]{Leiden University Medical Center}

\date{}
\maketitle


\begin{abstract}
\noindent

\textbf{Background:} Time-to-event data with multiple time scales are observed in many epidemiological and clinical studies. While models that allow for simultaneous consideration of multiple time scales for the hazard of an event have been proposed, their use is still not wide-spread in applied research. One reason for this might be the lack of convenient statistical software to estimate such models. Here we introduce the R-package \texttt{TwoTimeScales}. The package provides tools to estimate models for hazards that vary smoothly over two time scales, including proportional hazards models with such a two-dimensional baseline hazard. Extensions to competing risks models are implemented as well. Methodology is based on two-dimensional smoothing with $P$-splines.
\textbf{Results:} We demonstrate the features of the R-package by analysing a freely available dataset containing post-surgery follow-up data on patients with breast cancer. We present two examples, a proportional hazards regression and a competing risks problem. Besides estimation, we illustrate the plotting utilities of the package.  
\textbf{Conclusion:} The R-package \texttt{TwoTimeScales} can be easily used to fit flexible hazard models with two time scales, allowing new perspectives in the analysis of time-to-event data with multiple time scales.

\vspace{0.5cm}
\end{abstract}

\vskip 3ex

\section*{Background}\label{sec:background}

Time-to-event data are analyzed by means of survival models, where the hazard of an event evolves over the time scale of interest. In some applications, time-to-event data can involve more than one time scale. For example, in medical and epidemiological studies, time since disease onset and age of the patient (which is time since birth) may jointly determine the occurrence of an event, such as death or relapse after a cancer treatment. 

Recent advancements in the statistical literature have focused on flexible methods for estimating and representing hazard models over multiple time scales \citep{Batyrbekova:2022, Bower:2022, Carollo:2025, Carollo:2025b}, while previous research mostly proposed parametric models that rely on stronger assumptions \citep{Efron:2002, Iacobelli:2013, Rebora:2015}. Given the prevalence of time-to-event data with multiple time scales, the topic is of high methodological and applied importance.  

The hazard over two time scales can be modelled and estimated by two-dimensional $P$-splines \citep{Carollo:2025}. The approach can be extended to a proportional hazards (PH) model with a baseline hazard varying over two time scales. To efficiently estimate this model, array algorithms are employed \citep{Currie:2006, Eilers:2006}. Alternatively, the $P$-splines model can be represented as a mixed model and efficient algorithms for sparse matrices are available to obtain the estimated parameters \citep{Boer:2023}. The model presented in \citet{Carollo:2025}, and its extension to competing risks \citep{Carollo:2025b} are implemented in the R-package \texttt{TwoTimeScales} \citep{TwoTimeScales}.

Flexible parametric models (FPM) for survival analysis with multiple time scales have been implemented in Stata using the \texttt{stmt} package and command.
\citet{Bower:2022} provide a tutorial on how to use \texttt{stmt} to estimate a FPM with multiple time scales and present options for plotting the estimated hazard. To the best of our knowledge, the FPM with multiple time scales and the related Stata package, do not extend to competing risks models.

While there are currently a few R-packages that implement FPMs for the hazard, like \texttt{flexsurv} \citep{flexsurv} and \texttt{mexhaz} \citep{mexhaz}, or that allow for flexible estimation of piecewise exponential additive models, like \texttt{pammtools} \citep{pammtools,Bender:2018}, no R-package considers flexible models over multiple time scales simultaneously.

The focus of \texttt{TwoTimeScales} is on models for time-to-event data over two time scales, estimated through two-dimensional $P$-splines. The R-package \texttt{MortalitySmooth} \citep{Camarda:2012} implements two-dimensional $P$-splines for count data with offset, and it is particularly suited to smooth mortality surface at the population level, where individual survival data are aggregated to event counts and at-risk times. However, \texttt{MortalitySmooth} does not allow for regression models for the hazard. 

This paper describes the R-package \texttt{TwoTimeScales} and shows how to use the main functions of the package. We demonstrate its use by analysing follow-up data on patients with breast cancer who underwent surgery. The \texttt{rotterdam} dataset is available from the \texttt{survival} package \citep{survival-package} 
(see \verb|help(survival::rotterdam)|). The data were originally provided in \citet{Royston:2013}. We first analyse mortality rates after surgery by time since surgery and age with a regression model, without distinguishing between deaths before or after recurrence. Then, using the same data, we show how to fit a competing risks model to estimate the probabilities of recurrence of the cancer and of death without recurrence by time since surgery and age.  

In the next subsection we briefly review the two time scales $P$-splines model. We then introduce the \texttt{rotterdam} data in more detail, and discuss the data structure. The Implementation section is dedicated to the software. There we present the main functions of \texttt{TwoTimeScales}, describing their usage along with the most important arguments. We also provide code snippets demonstrating how to analyse the breast cancer data in the two examples. We report the results of the analyses, and compare our approach with the FPM approach implemented in Stata, in the Results section. Finally we conclude the paper by discussing how the proposed approach and package compare with other existing approaches, and with an outlook on future developments.

\subsection*{Hazard models with two time scales} \label{sec:model} 
In this section, we will briefly review the smooth hazard model with two time scales, implemented in the R-package \texttt{TwoTimeScales}. For a full exposition of the model, we refer to \citet{Carollo:2025}.

The two time scales are denoted by $t$ and $s$, where the origin of $s$ is later than the origin of $t$, so that $t>s$. The difference between the two time scales is fixed and we indicate with $u$; it is the value of $t$ when $s = 0$ \footnote{For example, if $t$ denotes the age of a patient and $s$ denotes the time since diagnosis, then $u=t-s$ is the age of the patient at diagnosis.}. The value of $u$, often indicating the timing of the intermediate event (for example, relapse of disease), differs between individuals. 

The estimand of interest is the hazard rate of an event over the two time scales. This hazard is defined as:
\begin{equation}
\label{eqch5:haz2D}
\lambda(t, s) = \lim _{\varepsilon \rightarrow 0+}\; 
\frac{P\left \{ \;\text{event} \in  L_\varepsilon (t,s)\;|\; \text{no event before } (t, s) \; \right \} }{\varepsilon} ,
\end{equation}
where we denote the line between $(t,s)$ and $(t+\varepsilon, s+\varepsilon)$, along which individuals advance, by $L_\varepsilon (t,s ) =\{ (t+\varepsilon ; s+\varepsilon) : \varepsilon \geq 0\}$.

The baseline hazard can be modified by some covariates of interest, in a proportional hazard fashion, so that $ \lambda(t,s; z) = \lambda_0(t,s)\exp(\beta^\top z)  ,$
where $\lambda_0(t,s)$ is the smooth baseline hazard over $t$ and $s$, $z$ is a vector of covariates, and $\beta$ is the vector of regression coefficients. Equivalently, we can express the model via $u=t-s$. Hence
\begin{equation}
	\lambda(t,s; z) = \lambda_0(t,s)\exp(\beta^\top z) \equiv \breve\lambda_0(u,s)\exp(\beta^\top z),
	\label{eqch5:ph}
\end{equation}
and we estimate $\breve\lambda_0(u,s)$ over $u>0, s>0$.

We estimate $\log(\breve\lambda(u,s))$ by two-dimensional $P$-splines. The baseline log-hazard $\breve\eta_0(u,s) = \log(\breve\lambda_0(u,s))$ is expressed as a linear combination of bivariate $B$-splines bases. 
In matrix formulation, the log-hazard for individual $i$ can be written as:
\begin{equation}\label{eqch5:PHmatrix}
E_i = B_uAB_s^\top + x_i^\top\beta , 
\end{equation}
where the matrix $A = [\alpha_{lm}]$, of dimensions $c_u \times c_s$, contains the $c_uc_s$ coefficients for the $B$-splines.
The model is fitted via penalized Poisson regression, and the estimates are obtained through the Iterative Weighted Least Squares (IWLS) algorithm (see \citet{Carollo:2025}).

In Equation~\eqref{eqch5:PHmatrix}, the matrix $A$ contains all $B$-splines coefficients. A penalty matrix penalizes differences in adjacent coefficients of $A$, to ensure smooth results. The rows and columns of the matrix $A$ are penalized separately by choosing two different smoothing parameters $\varrho_u$ and $\varrho_s$ (also known as anisotropic smoothing). 

Once the optimal pair of smoothing parameters is selected the matrix of estimated parameters for the log-hazard, together with their associated standard errors is returned. These estimates can be used to evaluate the (log-)hazard surface at a finer grid of points, to obtain a more detailed surface. From the standard errors of the $\alpha$ coefficients, corresponding uncertainty measures are obtained for the (log-)hazard by using the delta-method, as shown in \citet{Carollo:2025}.

The extension of the model to competing risks has been discussed in \citet{Carollo:2025b}. In short, the cause-specific hazards are fitted as separate models for each event type. The cumulative cause-specific hazards are obtained by numerical integration along the $s$-scale, from which the overall survival function is obtained. Finally, the cumulative incidences are then obtained by multiplying the cause-specific hazards with the overall survival function. Uncertainty measures for the cumulative incidence functions are obtained by non-parametric bootstrapping.

\subsection*{Illustration: Rotterdam breast cancer data}
\label{sec:breastdata}
\begin{figure}
\centering
\includegraphics[width=.8\textwidth]{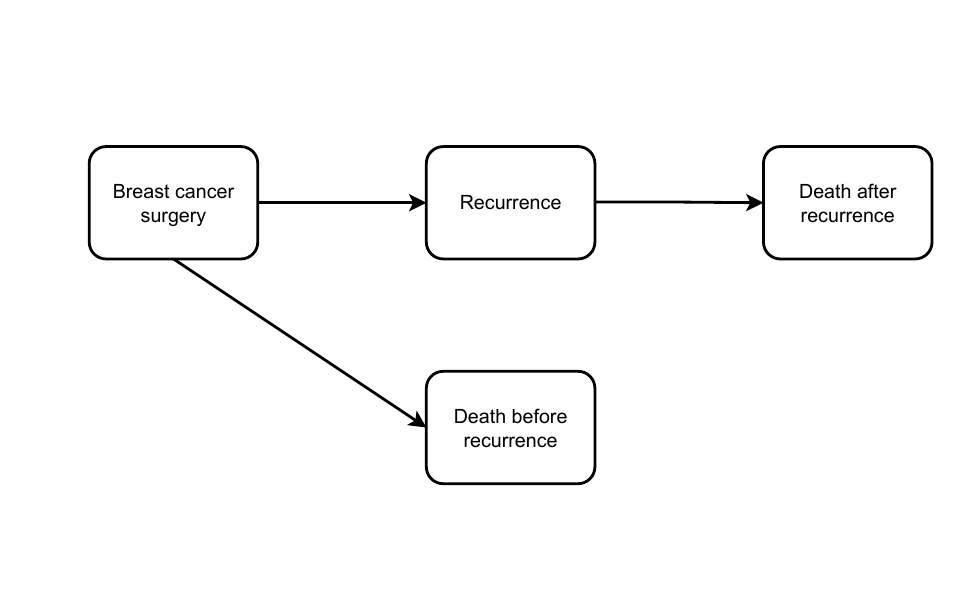}
\caption{Multistate model of the \texttt{rotterdam} breast cancer data.}
\label{figch5:bcstates}
\end{figure}

The \texttt{rotterdam} dataset contains follow-up data on 2,982 patients with primary, operable and invasive breast cancer selected from the Rotterdam tumour bank, who received surgery (modified mastectomy or breast conserving surgery) and were followed-up until recurrence of the cancer and/or death, or end of follow-up.
After surgery, patients could either experience a recurrence of the cancer as a local recurrence or a distant metastasis, or they could die without recurrence. Hence, recurrence of the cancer and death without recurrence are two competing events after surgery. Among individuals who have experienced recurrence of the cancer, some will eventually die during follow-up. The data can be represented by the multistate process in Figure~\ref{figch5:bcstates}. 

In the first illustrative example, we will analyse time to death (irrespective of recurrence) by \textit{time since surgery} and \textit{age} for all $2,982$ individuals. The same data are analysed in the paper by \citet{Bower:2022}. We will briefly compare the results of our analyses with those obtained by Bower and colleagues using their Stata package in the Result section.

The following code snippet shows how to load the data in R, rescale the time variables in years, and create additional time variables such as the ages at recurrence and death. In the description of \texttt{rotterdam} dataset in \texttt{survival}, a warning is given regarding 43 individuals whose time to recurrence is censored at an earlier time than the last follow-up time, which for these individuals is the time of death. It is not possible to definitely conclude that these patients died without recurrence, so we will censor these patients at the time of censoring for recurrence.

\begin{lstlisting}[language = R]
library("survival")                          # for dataset rotterdam
d <- rotterdam                               # dataset

# Extract time variables
d$rtimey <- d$rtime/365.25                   # time from surgery to 
                                             # recurrence in years
d$dtimey <- d$dtime/365.25                   # time  from surgery to
                                             # death in years

# Extract age variables
# d$age is the age at surgery
d$rage <- d$age + d$rtimey                   # age at recurence
d$dage <- d$age + d$dtimey                   # age at death

# Address concern about the 43 individuals with censoring time for recurrence  smaller than death times
d[d$rtime < d$dtime & d$recur == 0, ]$death <- 0
d[d$rtime < d$dtime & d$recur == 0, ]$dtime <-  
    d[d$rtime < d$dtime & d$recur == 0, ]$rtime
\end{lstlisting}

In the second example, we consider recurrence and death without recurrence as competing events. Again, time since surgery and age of the patient are the two time scales. For this second analysis, we create two additional variables - the first is the time in years to the first event after surgery, and the second is a variable that distinguishes between recurrence or death without recurrence for the first observed event after surgery, as shown by the following code snippet.

\begin{lstlisting}[language = R]
# Extract time to first event
d$fetimey <- pmin(d$rtimey, d$dtimey)

# Code type of first event
d$first_event <- ifelse(d$recur == 1, "recurrence", NA)
d$first_event <- ifelse(d$recur == 0 & d$death == 1, 
                        "death", d$first_event)
d[is.na(d$first_event),]$first_event <- "censored"
\end{lstlisting}

\section*{Implementation}
\subsection*{Software: \texttt{TwoTimeScales}}\label{sec:TwoTimeScales}
The R-package \texttt{TwoTimeScales} offers functions for analysing time-to-event data with two time scales by means of the two-dimensional $P$-splines model reviewed in the previous section. \texttt{TwoTimeScales} can be installed directly for The Comprehensive R Archive Network (CRAN) via \verb|install.packages(`TwoTimeScales')| and loaded into the \\ R-environment using \verb|library(TwoTimeScales)|.

\begin{figure}
\centering
\includegraphics[width=\textwidth]{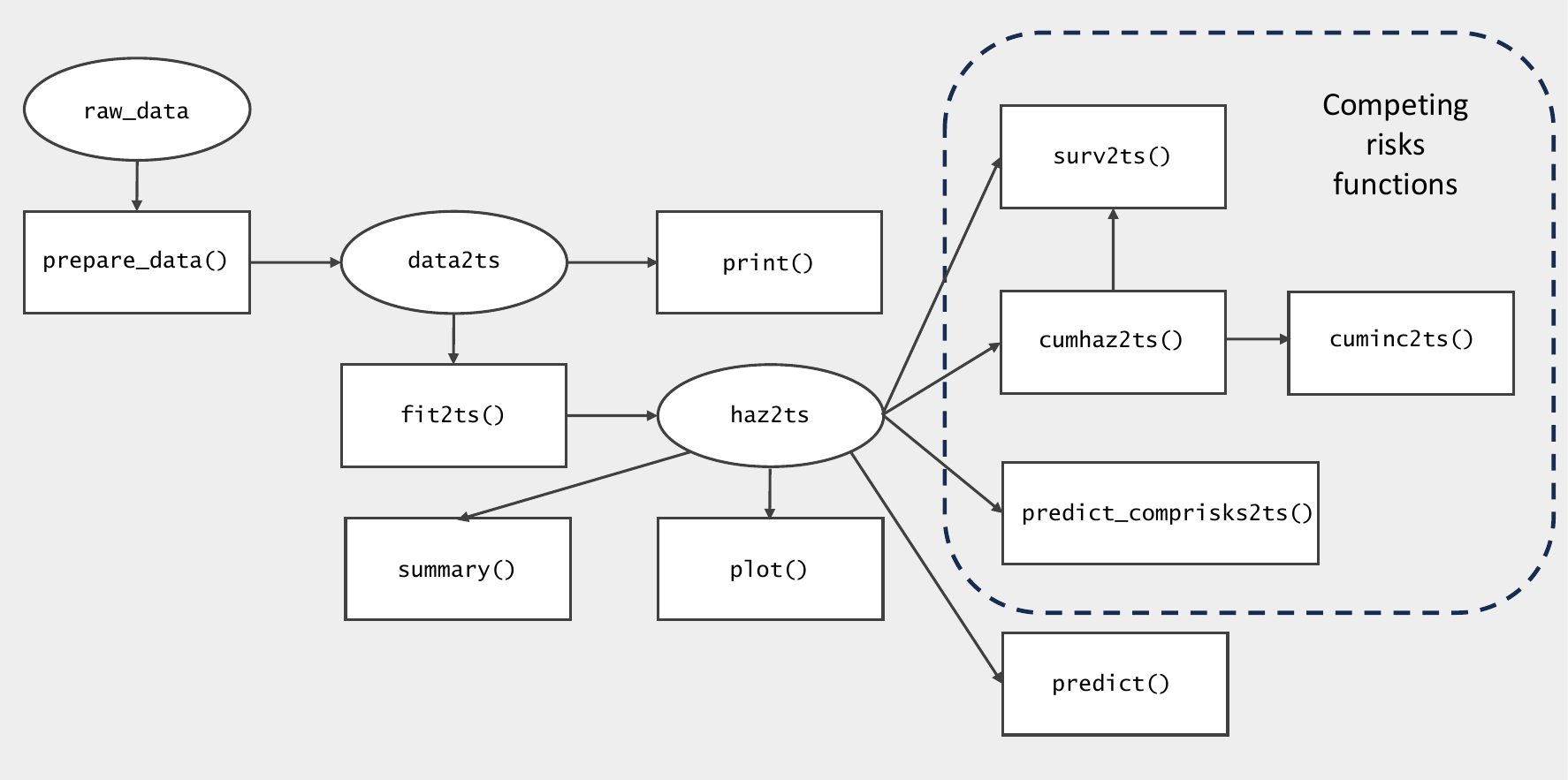}
\caption{Flowchart illustrating the workflow of \texttt{TwoTimeScales}. R-objects, indicated by their class when appropriate, are in ovals, while functions are in rectangles. Functions to extend the analysis to a competing risks model are enclosed within the dashed box.}
\label{figch5:flowchart}
\end{figure}

The main functionalities of the package allow for data preparation, model estimation, visualization of the results and predictions with arbitrary values of covariates and time scales. A general workflow is presented in Figure~\ref{figch5:flowchart}. For the analysis of competing risks over two time scales, the package provides additional functions to estimate cumulative quantities, such as the cumulative hazard, survival probabilities and the cumulative incidence functions, as presented in \citet{Carollo:2025b}.
 
In the following we will present in detail the main functions with which the user will commonly interact, namely \texttt{prepare\_data()}, \texttt{fit2ts()} and the methods \texttt{plot()} and \texttt{predict()} for objects of class \texttt{haz2ts}. We show how to use these functions to analyse the breast cancer data introduced in the Background section, separately for the two examples. The main results of our analysis are presented in the next section. All R-code can be found in an online repository \url{https://github.com/AngelaCar/RotterdamBC_analyses}. 

\subsubsection*{Preparing the data}
Most time-to-event datasets are provided in a wide format, that is one row per subject in the data. They usually include a time variable, possibly recorded in terms of starting time and stopping time, and an event indicator. In the simplest case, there is only one type of event, so that this event indicator can only assume values 0 and 1. Covariates are usually provided as additional columns in the dataset. To apply the $P$-splines model for the hazard we need to bin the individual data into arrays of exposure times and event indicators.

Data preparation involves two steps which are both performed by the function \texttt{prepare\_data()}: building a grid of bins and binning the individual data into these bins. The function \texttt{prepare\_data()} requires as input a \texttt{data.frame}, where to find the of values of $u$, the (exit) values for $s$ and the event indicators. Additionally, the desired width of the bins over the $s$ axis needs to be passed to the argument \texttt{ds}. The function \texttt{prepare\_data()} returns an object of class \texttt{data2ts}, and the print method is available for this class of objects, returning some basic info about the binned data. Table~\ref{tab:preparedata} summarizes the most important information about the parameters of the function \texttt{prepare\_data()}. 
\vskip 2ex
\begin{table}[]
    \centering
    \begin{tabular}{lp{6cm}cc}
        \textbf{argument} &  & \textbf{default value} & \textbf{required} \\
         \hline
         \texttt{data} & A data.frame with the individual data & none & no \\
         \texttt{u} & fixed difference between the time scales & none & no  \\
         \texttt{s\_in} & entry times over $s$ & none & no\\
         \texttt{s\_out} & exit times over $s$ & none & yes\\
         \texttt{events} & event indicators & none & yes \\
         \texttt{ds} & width of bins over $s$ & none & yes \\
         \texttt{du} & width of bins over $u$ & \verb|du=ds| & no\\
         \texttt{min\_u} & minimum value for $u$ & none & no \\
         \texttt{max\_u} & maximum value for $u$ & none & no \\
         \texttt{min\_s} & minimum value for $s$ & none & no \\
         \texttt{max\_s} & maximum value for $s$ & none & no \\
         \texttt{individual} & logical value indicating whether individual matrices of event counts and exposure times are needed & \verb|FALSE| & no \\ 
         \texttt{covs} & the names of the covariates, or a matrix of covariates values extracted from the original data.frame & none & no
    \end{tabular}
    \caption{Main arguments of the function \texttt{prepare\_data()}.}
    \label{tab:preparedata}
\end{table}

\noindent \textit{Example 1: All deaths (irrespective of recurrence)}\\
The following code snippet shows how to load the \texttt{TwoTimeScales} package and how to use the function \texttt{prepare\_data()} to prepare the data for the analysis of all deaths. Here, and in what follows, $u$ is \textit{age at surgery}, and $s$ is \textit{time since surgery}. We choose bins of size \verb|du=1| for the age at surgery (ranging from 24 to 90), while we choose \verb|ds=.5| for the time since surgery, that spans about 19 years. The different bins widths over the two axes reflect the expectation that the hazard of death might change more rapidly over time since surgery than over age at surgery. We include also the variable \texttt{grade} which measures tumour's differentiation grade (first transforming it into a factor). The same code for the second example, is provided in the supplementary material.

\begin{lstlisting}[language = R]
# install.packages("TwoTimeScales")          # install from CRAN
library("TwoTimeScales")
d$grade <- as.factor(d$grade)
death_cov <- prepare_data(data = d,
                          u = "age",         # age at surgery
                          s_out = "dtimey",  # time since surgery 
                          events = "death",
                          ds = .5, du = 1,
                          min_u = 24, min_s = 0,
                          individual = TRUE,
                          covs = c("grade")) # covariate
\end{lstlisting}

The data are now in the three-dimensional arrays of dimension $n_u \times n_s \times n$, as the matrices $R$ and $Y$ for the binned data are calculated for each individual, and a regression matrix $Z$ of dimension $n \times p$. Here $n_u$ and $n_s$ are the number of bins constructed over the $u$ and $s$ axes respectively, $n$ is the number of individuals in the data, and $p$ the number of covariates.
The print method applied to objects of class \texttt{data2ts} returns, along with some general information about the structure of the object, the range of values for the two time axes, the number of bins over each axis and the names of the dummy variables coded within \texttt{prepare\_data()}.
\begin{lstlisting}[style=Routput, caption={R output function \texttt{print(death\_cov)}}]
> print(death_cov)
An object of class 'data2ts'

Data:
List of 2
 $ bins   :List of 6
 $ bindata:List of 3
 - attr(*, "class")= chr "data2ts"
NULL

Range covered by the bins: 
$bins_u
[1] 24 90

$bins_s
[1]  0.0 19.5


Number of bins: 
$nu
[1] 66

$ns
[1] 39


Overview of the binned data:
Total exposure time: 21194.75
Total number of events: 1229
Covariates:
[1] "grade_3"
\end{lstlisting}

\subsubsection*{Fitting the model}
The function \texttt{fit2ts()} performs the actual estimation of the model. Its input \texttt{data2ts} was created by \texttt{prepare\_data()}, that is an object of class \texttt{data2ts}. It can include a design matrix for the covariates part of the PH model and accordingly the function will estimate a full PH model. Alternatively, the user can provide each element separately, that is a list of bins (\texttt{bins}), the arrays of exposure times (\texttt{R}) and event count (\texttt{Y}), and a regression matrix (\texttt{Z}).

If not specified otherwise, the function \texttt{fit2ts()} uses default settings for the degree of $B$-splines (\texttt{bdeg=3}), the order of the difference penalty (\texttt{pord=2}) as well as the number of segments for the $B$-splines (default is 10). The default method to determine the optimal values of the smoothing parameters $\varrho_u$ and $\varrho_s$ is numerical optimization of $\text{AIC}(\varrho_u, \varrho_s)$. The user can change the default settings, for details see the Appendix.

The object returned by \texttt{fit2ts()} is of class \texttt{haz2ts}, which is a list containing, among others, the parameter estimates (including standard errors), optimal smoothing parameters, AIC/BIC values, the full variance-covariance matrix of the estimated coefficients, and the effective dimension of the estimated model. The summary, plot and predict methods are available for objects of class \texttt{haz2ts}.

\vskip 2ex
\noindent \textit{Example 1: all deaths} \\
Here, we illustrate how to perform estimation of the two time scale PH model for the breast cancer data prepared in the previous step. A summary of the model shows the number of events, the number of bins built in each of the two axes and the number of $B$-splines parameters on each of the axis. The optimal smoothing parameters selected are shown, both on a $\log_{10}-$scale and in their original scale, as well as the effective dimension (ED). If a PH model is estimated, a table with the covariate effects, their associated standard errors, the hazard ratios (HR) and the 95\% confidence intervals of the HR, is shown. Finally, the AIC and BIC of the fitted model are reported.
\begin{lstlisting}[language = R]
mod_deathcov <- fit2ts(data2ts = death_cov,
                       Bbases_spec = list(bdeg = 3,
                                          nseg_u = 12,
                                          min_u = 24,
                                          nseg_s = 7,
                                          min_s = 0),
                       pord = 2,
                       lrho = c(1,-2),
                       optim_method = "ucminf",
                       optim_criterion = "bic")
summary(mod_deathcov)
\end{lstlisting}

\begin{lstlisting}[style=Routput, caption={R output function \texttt{summary(mod\_deathcov)}}]
> summary(mod_deathcov)
Number of events =  1229 
Model specifications:
  nu =  66 
  ns =  39 
  cu =  15 
  cs =  10 

Optimal smoothing: 
  log10(rho_u) =  2.580813 
  log10(rho_s) =  -0.5666426 
  rho_u =  380.902 
  rho_s =  0.2712423 

             beta   se(beta) exp(beta) lower .95 upper.95
grade_3 0.5083305 0.07134667  1.662513  1.430028 1.894998


Model diagnostics: 
  AIC =  11001.32 
  BIC =  11101.93 
  ED =  11.57828
\end{lstlisting}
The computation time for fitting this model was 4m64s on a laptop with Intel(R) Core(TM) Ultra 5 135H, 16GB RAM and Windows operating system. 
The code to estimate the two cause-specific hazards in the second example is very similar, so it is omitted here and available in the online repository.

\subsubsection*{Plotting the estimated hazard}
The method \texttt{plot()}, for objects of class \texttt{haz2ts} offers different possibilities for visualizing the results of the hazard model with two time scales. The user can choose between plotting the hazard surface \texttt{which\_plot = "hazard"}, plotting the standard error surface \texttt{which\_plot = "SE"}, plotting cross-sections of the surface (also called `slices') at specific values of one of the two time dimensions \texttt{which\_plot = "slices"}, plotting the survival function \texttt{which\_plot = "survival"} or the cumulative hazard \texttt{which\_plot = "cumhaz"}, and finally, for PH models with covariate effects, it is possible to plot the estimated $\beta$ coefficients, or the corresponding hazard ratios, with their confidence intervals with \texttt{which\_plot = "covariates"}.

The default plot is the hazard surface over the $(u,s)$-plane, with white contour lines to mark levels of the hazard. With the default option, the surface will be evaluated at the grid-points of the two-dimensional bins that were used for binning of the data and estimation. Should one want a finer plotting grid (and hence a smoother surface plot) the argument \texttt{plot\_grid} allows to specify a new grid. In the example below the grid points for the plot will be 0.2 and 0.1 years apart for $u$ and $s$ respectively, while the bin were separated by 1 and 0.5 years in the binned data.

A number of optional graphical parameters and plotting options can be passed to the argument \texttt{plot\_options}. %
We provide an example in the following code snippet, where we illustrate some of the options, such as axis labels and plot titles. Other examples of the usage of the \texttt{plot} method can be found in the dedicated vignette of the package.
The option \texttt{original} selects whether the hazard is plotted over the $(t,s)$-plane (\texttt{original = TRUE}) or over the $(u,s)$-plane (\texttt{original = FALSE}); the latter is the default setting. The penalties allow to extend the bivariate hazard into areas where no data were observed, however, extrapolation beyond the data in general has to be handled cautiously (see discussion in \cite{Carollo:2025}). The plot options \texttt{tmax} and \texttt{cut\_extrapolated} allow to limit the plotted hazard hazard surface to areas supported by data, and here we limit it to current age 90, see the example below.

A plot of the standard error surface for the two-dimensional hazard can be obtained by setting the argument \texttt{which\_plot = "SE"}. The same arguments that control plotting options for the hazard surface can be used to control those for the SEs surface, for example plotting the SEs of the hazard or of the log-hazard values. 

\begin{figure}[!h]
    \centering
    \begin{subfigure}[t]{0.48\textwidth}
        \centering
        \includegraphics[width=\textwidth]{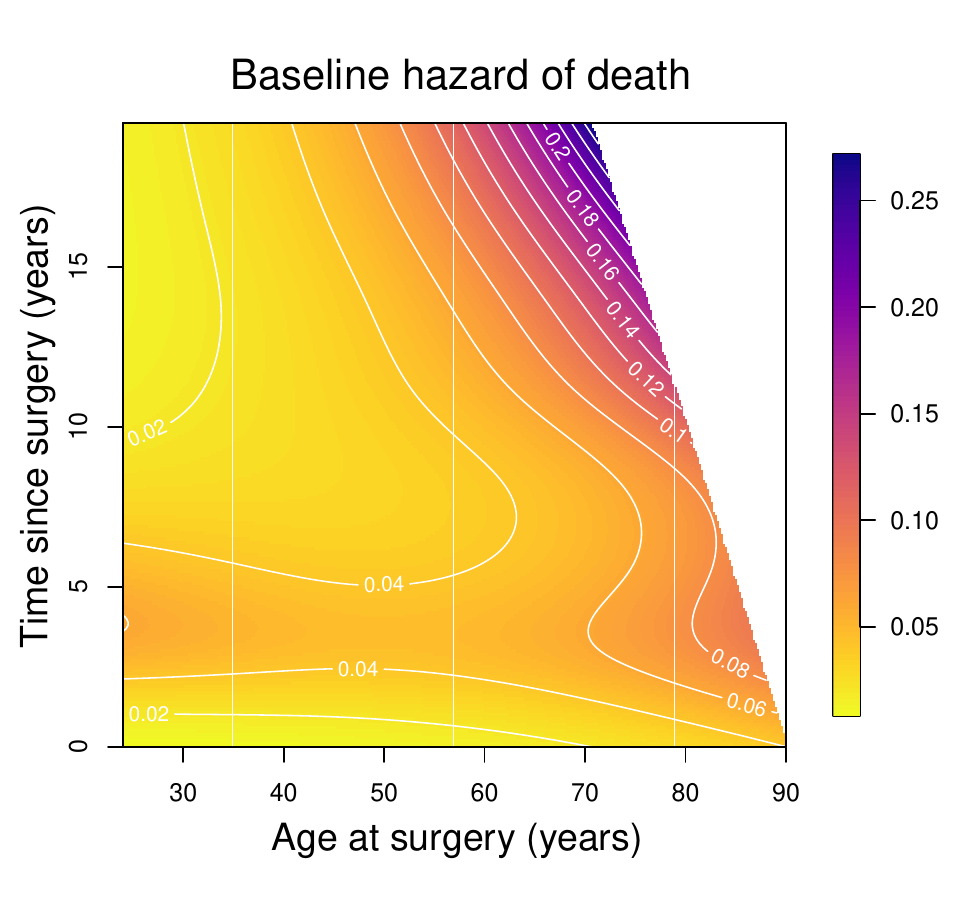}
        \caption{$\hat{\lambda}_0(u,s)$.}
    \end{subfigure}%
    ~~~ 
    \begin{subfigure}[t]{0.48\textwidth}
        \centering
    	\includegraphics[width=\textwidth]{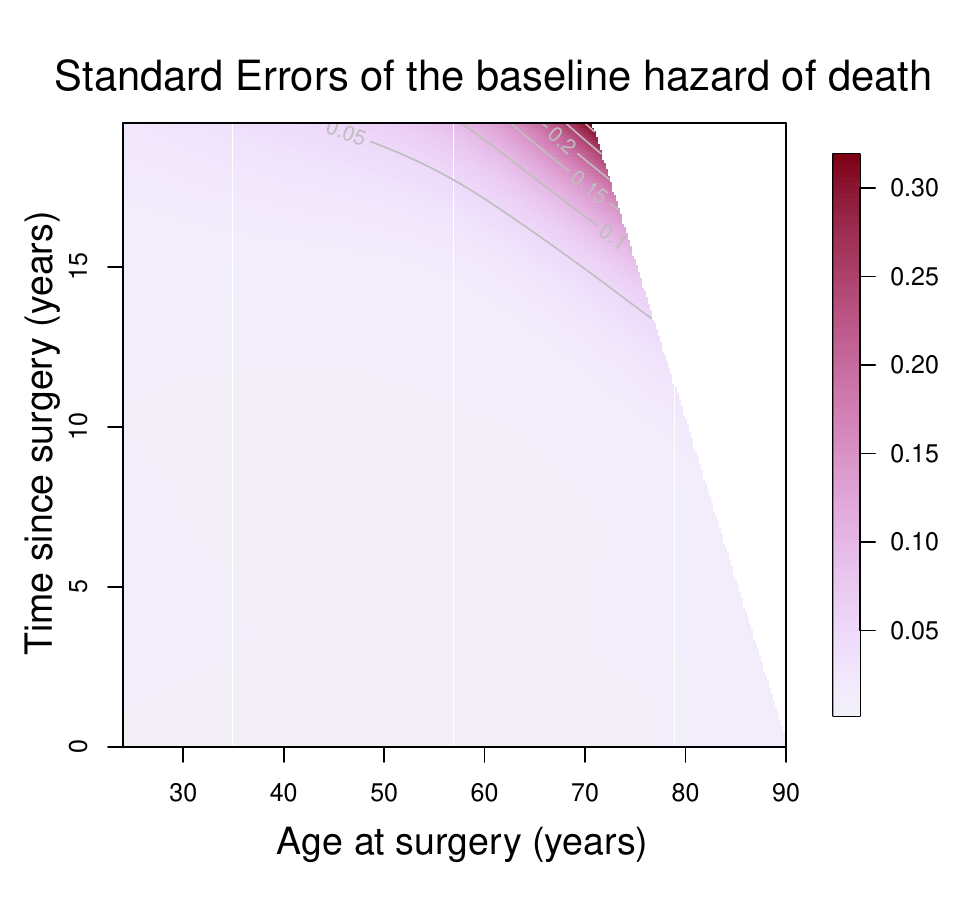}
		\caption{$\text{SE}(\hat{\lambda}_0(u,s))$.}
    \end{subfigure}
\caption{Baseline hazard of death in the breast cancer data, represented on the $(u,s)$-plane (left) and on associated SE surface (right).}
\label{figch5:2hazards}
\end{figure}

The following code produces the two plots in Figure~\ref{figch5:2hazards}, that is the estimated baseline hazard of death over age at surgery and time since surgery, the $(u,s)-$plane on the left panel, and its associated SE surface on the right panel.

\begin{lstlisting}[language = R]
# Hazard on the (u,s)-plane
plot(mod_deathcov,
     plot_grid = list(c(umin = 24, umax = 90,
                        du = .2),
                      c(smin = 0, smax = 19.5, 
                        ds = .1)),
     plot_options = list(n_shades = 100,
                         tmax = 90,
                         cut_extrapolated = TRUE,
                         main = "Baseline hazard of death",
                         xlab = "Age at surgery (years)",
                         ylab = "Time since surgery (years)",
                         xlim = c(24, 90)))
# Associated SEs surface over (u,s)-plane
plot(mod_deathcov,
     which_plot = "SE",
     plot_grid = list(c(umin = 24, umax = 90,
                        du = .2),
                      c(smin = 0, smax = 19.5, 
                        ds = .1)),
     plot_options = list(n_shades = 100,
                         tmax = 90,
                         cut_extrapolated = T,
                         main = "Standard Errors of the baseline hazard of death",
                         xlab = "Age at surgery (years)",
                         ylab = "Time since surgery (years)",
                         xlim = c(24, 90)))
\end{lstlisting}

Plotting cross-sections of the two-dimensional hazard as curves over one of the time scales is another informative representation. For example, the hazard surface is plotted as a series of one-dimensional hazard curves over time since recurrence $s$ for selected values of time at recurrence $u$. Such a plot is obtained by setting the argument \texttt{which\_plot = "slices"}. Additionally, the user can specify the direction where the cross-sections are to be obtained, \texttt{direction = "u"} for hazard curves over $s$ and vice-versa (that is \texttt{direction = "s"}, for cross-sections of the hazards at specific values of $u$), and the values for the cutting points. By default, these cross-sections of the hazard will be plotted on a grey color-scale, but the user can specify any desired color palette in the argument \texttt{col\_palette}. 

It is possible to plot the survival function and the cumulative hazard both as a two-dimensional surface over the $(u,s)-$ or the $(t,s)-$plane, or as one-dimensional curves over $s$ for selected values of the entry time $u$, by selecting the corresponding plot type in \texttt{which\_plot} and, if plotting one-dimensional curves, a vector of cutting points should be passed to the argument \texttt{where\_slices} and the plotting option \texttt{surv\_slices} or \texttt{cumhaz\_slices} should be set to \texttt{TRUE}. 

\subsubsection*{Predicting hazard values}
A prediction method is implemented for objects of class \texttt{haz2ts}. Given a fitted model and a \texttt{data.frame} with values of the variables \texttt{u}, \texttt{s} and if desired the values of the covariates in the model, the user can obtain predicted hazard values, cumulative hazard values and survival probabilities for each pair of $(u,s)$-values and specific values of the covariates. If the original dataset is passed to the function, then predicted quantities are obtained for each of the observed $(u,s)$ data points. Alternatively, the user can provide a new dataset with values of $u$ and $s$ within the range of the observed variables.
An example of its usage is provided in the next code-snippet, and the full predictions are shown in Figure~\ref{figch5:predictions}.
\begin{lstlisting}[language = R]
newdata <- as.data.frame(expand.grid("age" = c(40, 50, 60), 
                                     "dtimey" = seq(0, 15, by = .1),
                                     "grade_3" = 1))
newdata$t <- newdata$age + newdata$dtimey
newdata <- subset(newdata, t <= 70)
predicted_haz <- predict(mod_deathcov,
                         newdata = newdata,
                         u = "age", s = "dtimey", z = "grade_3")
round(head(predicted_haz), 3)                         
\end{lstlisting}

\begin{lstlisting}[style=Routput, caption={R output function \texttt{predict(mod\_deathcov)}}]
> round(head(predicted_haz),3)
  age dtimey grade_3    t hazard cumhazard se_hazard
1  40    0.0       1 40.0  0.017     0.002     0.003
2  50    0.0       1 50.0  0.020     0.002     0.003
3  60    0.0       1 60.0  0.025     0.002     0.003
4  40    0.1       1 40.1  0.018     0.004     0.003
5  50    0.1       1 50.1  0.021     0.004     0.003
6  60    0.1       1 60.1  0.026     0.005     0.003
  survival basehazard se_basehazard
1    0.998      0.010         0.002
2    0.998      0.012         0.002
3    0.998      0.015         0.002
4    0.996      0.011         0.002
5    0.996      0.013         0.002
6    0.995      0.016         0.002
\end{lstlisting}

\subsubsection*{Calculating competing risks quantities}
\label{sec:cif}
Here we will show how to use \texttt{TwoTimeScales} to calculate the cumulative incidence functions over two time scales, using the breast cancer data example discussed in the Background. 
After receiving surgery for invasive breast cancer women were at risk of experiencing a recurrence of the cancer, or dying without recurrence. We analyse the transitions to these two competing events over age of the woman and time since surgery in a competing risks model. 

The first step is to fit cause-specific hazards for each competing event, keeping the specification of the data and model (that is, number of bins, number of $B$-splines, and order of the penalty) the same among all models. From the fitted cause-specific hazards we can obtain estimates of the cumulated quantities as described in the Background section.

In the following code snippet we show how to obtain the cause-specific cumulative incidences for the breast cancer. We will obtain the estimated incidence of recurrence and of death without recurrence by age of the women and time since surgery. The code for the full analysis of these data is available in the repository, while here we only focus on the steps from the fitted cause-specific models to the cumulative incidence functions.

The function \texttt{cumhaz2ts()} requires a fitted model, that is an object of class \texttt{haz2ts}, and the specifications for a grid of values where the estimated hazard can be evaluated and then cumulated over the $s$-axis. This is optional, and if none is provided, then the same grid used for estimation will be also used for this cumulation. The object produced by \texttt{cumhaz2ts()} is a list with estimated values for the estimated hazard surface and the estimated cumulated hazard surface. The optional parameter \texttt{cause} is a string which contains a short name for the event type.

The function \texttt{surv2ts()} can be called to obtain an estimate of the two-dimensional survival function from a fitted model. Similarly to \texttt{cumhaz2ts()} the user can pass as arguments a fitted model and an optional list with parameters for a new plotting grid. Alternatively, \texttt{surv2ts()} can also be provided with a list of cause-specific integrated hazards obtained from \texttt{cumhaz2ts()} (\texttt{cumhaz}), and a vector with short names for each cause. 

Finally, the function  \texttt{cuminc2ts()} requires as inputs a list containing the cause-specific hazards \texttt{haz} and the distance between grid-points on the $s$ axis \texttt{ds}. Optionally, a vector with short names of each cause can be passed to the argument \texttt{cause}. The function \texttt{cuminc2ts()} returns a list with the estimated cumulative incidence functions over $u$ and $s$.

\begin{lstlisting}[language = R]
H_rec <- cumhaz2ts(mod_rec,
                   plot_grid = list(c(umin = 24, umax = 90, 
                                      du = .2),                               
                                    c(smin = 0, smax = 19.5,  
                                      ds = .1)),
                   cause = "recurrence")
H_death <- cumhaz2ts(mod_death,
                     plot_grid = list(c(umin = 24, umax = 90, 
                                        du = .5), 
                                      c(smin = 0, smax = 19.5, 
                                        ds = .1)),
                     cause = "death")
cif <- cuminc2ts(haz = list(H_rec$Haz$hazard, H_death$Haz$hazard),
                 ds = .1,
                 cause = c("recurrence", "death"))
\end{lstlisting}

\section*{Results}\label{sec:results}
\subsubsection*{Example 1: all deaths irrespective of recurrence}
In this section we present additional results of the two time scales hazard model for the transition from surgery to death (with or without recurrence), for the \texttt{rotterdam} breast cancer data. The time scales are $t$, \textit{current age}, and $s$, \textit{time since surgery}. The \textit{age at surgery} $u$, is obtained as difference between these two time scales.

We estimate a PH model with the covariate \texttt{grade}, we specify 15 marginal $B$-splines over the $u$ axis and 10 $B$-splines over the $s$ axis, and select the optimal smoothing parameters by a numerical optimization of the BIC as a function of the smoothing parameters. The fitted model has ED = 11.58, and the optimal smoothing parameters selected are $\log_{10}(\varrho_u) = 2.58$ and $\log_{10}(\varrho_s) = -0.57$. The estimated effect of the covariate \texttt{grade} is 0.51, that translates to $HR=1.66$ (95\% CI: $1.43, 1.89$). Women with breast cancer of grade 3 have about 66\% higher risk of death than women with a grade 2 breast cancer.

The baseline hazard is graphically depicted in Figure~\ref{figch5:2hazards}, over the transformed $(u,s)$-plane. From Figure~\ref{figch5:2hazards} the complex shape of the baseline hazard of death after surgery emerges well. The hazard of death (with or without recurrence) increases in the first 4 to 5 years almost homogeneously for all ages at surgery, and decreases afterwards but only for women who had surgery before age 5. For women who had surgery after age 50 the hazard of death 5 or more years after surgery either stays constant or increases even further.

\begin{figure}[!h]
   \begin{subfigure}[t]{0.48\textwidth}
        \centering
        \includegraphics[width=\textwidth]{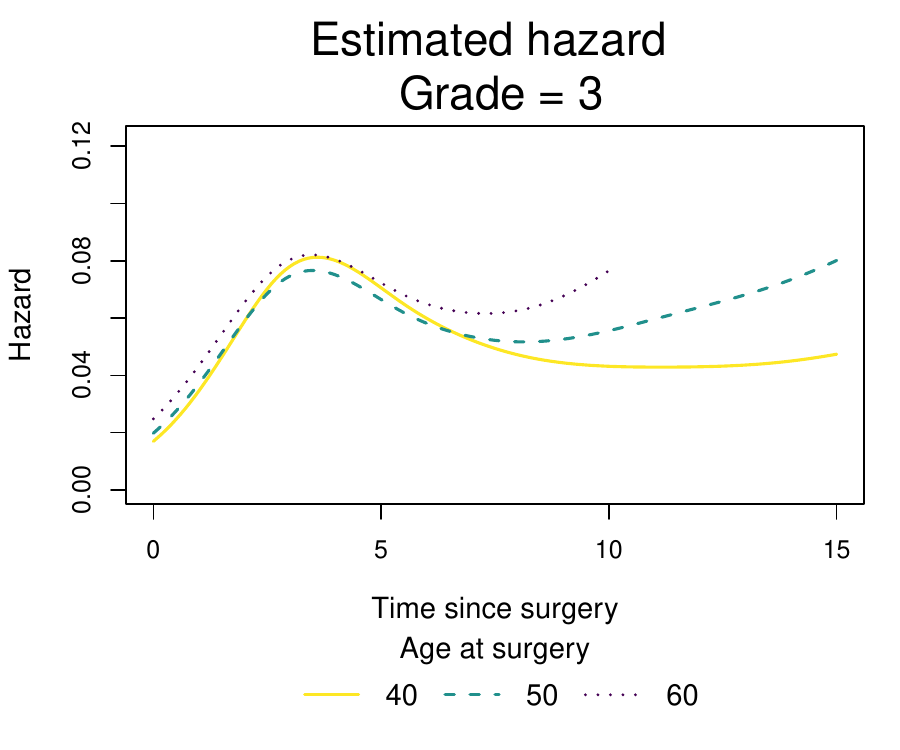}
   \end{subfigure}%
    ~~~ 
    \begin{subfigure}[t]{0.48\textwidth}
        \centering
        \includegraphics[width=\textwidth]{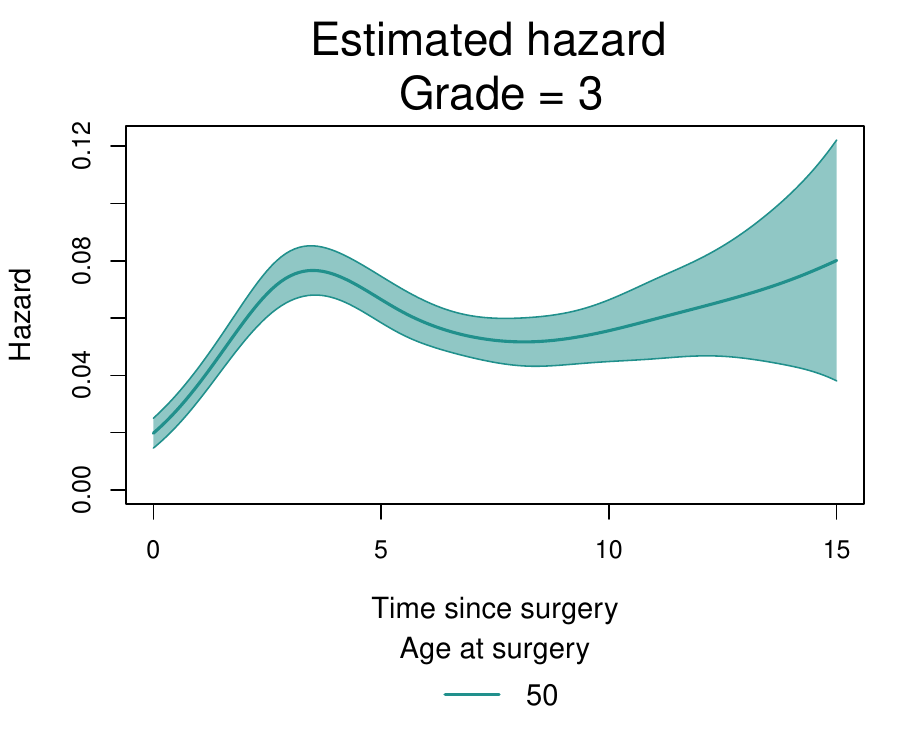}
    \end{subfigure}%
\caption{Left: Mortality rates after surgery, for grade 3 breast cancer, over time since surgery and selected ages at surgery. Right: Mortality rate after surgery for a woman with breast cancer grade 3 who had surgery at age 50, with CIs.}
    \label{figch5:predictions}
\end{figure}

The two plots in Figure~\ref{figch5:predictions} are directly comparable to Figure 2(a) and (b) in \citet{Bower:2022}. Our results are almost identical to those obtained using the FPM implemented in the Stata package. Other plots over time since surgery for selected values of current age or over current age for selected time since surgery (as in Figure 2(c) and (d) of \citet{Bower:2022}) are easily obtainable, and here omitted.

\subsubsection*{Example 2: competing risks after surgery}
\begin{figure}[!h]
\centering
\includegraphics[width=\textwidth]{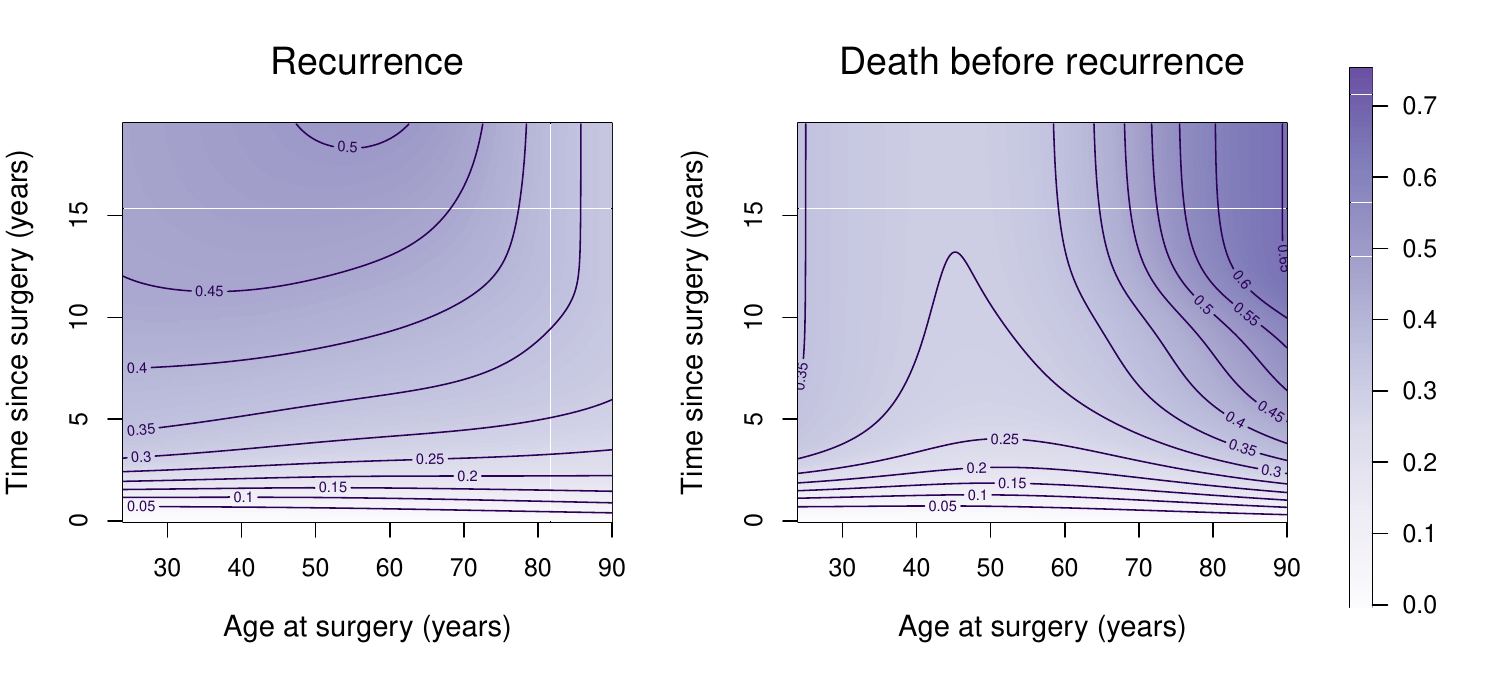}
\caption{Cumulative incidence functions of recurrence (left panel) and death before recurrence (right panel) by age at surgery and time since surgery.}
\label{figch5:CIF}
\end{figure}

The estimated cumulative incidence of recurrence and of death before recurrence over age at surgery (horizontal axis) and time since surgery (vertical axis) are shown in Figure~\ref{figch5:CIF} with a common legend. 

The probability of experiencing a recurrence of the cancer increases with time since surgery, similarly over age at surgery. In contrast, the probability of dying without recurrence increases in a much faster way, over time since surgery, for women who had surgery at age 50 or older.

\section*{Discussion}
\label{sec:discussion}
We presented the R-package \texttt{TwoTimeScales} which provides tools for the analysis of time-to-event data with two time scales. The users will usually interact with four main functions, \texttt{prepare\_data()}, \texttt{fit2ts()}, \texttt{predict()}, and \texttt{plot()}, that perform data preparation, fitting of the two-dimensional $P$-spline model for the hazard, prediction tasks, and plotting the resulting surfaces, respectively.
In case of competing risks models, additional functions are provided that allow the ultimate goal of estimating probabilities of cause-specific events over two time scales (CIFs).
 
\texttt{TwoTimeScales} implements a fast version of the GLAM algorithm \citep{Currie:2006} to fit two-dimensional $P$-spline models to Poisson data with offset in a PH regression setting. Alternatively, functions from the \texttt{LMMsolver} R-package \citep{Boer:2023} are integrated that enable even faster computations by exploiting the sparse matrix structure and the link to linear mixed models (see \citet{Carollo:2025} for details).

\texttt{TwoTimeScales} also provides functions to estimate smooth hazard models over one time scale with $P$-splines, and we provide a tutorial in a vignette of the package. 

While the approach implemented in \texttt{TwoTimeScales} explicitly models the simultaneous effect of two time scales and their interaction on the hazard of events, in some circumstances a simpler model might suffice. \citet{Carollo:2025c} compare three models of increasing complexity: a log-additive model, a varying coefficients model, and the fully interactive model also presented in this paper. Functions to estimate these simpler models, also using $P$-splines are implemented in \texttt{TwoTimeScales}, as well as a function which selects the best fitting model among a list of models, using either AIC or BIC criteria. 

In this article we have analysed the same dataset analysed in \citet{Bower:2022}. The results of the model estimated here with \texttt{TwoTimeScales} are virtually identical to those produced using the FPM and Stata package, therefore affirming \texttt{TwoTimeScales} as a valid alternative to fit smooth hazard models over two time scales for R-users. Additionally, \texttt{TwoTimeScales} allows to extend the analyses to settings with competing events. To the best of our knowledge, competing risks models are not yet implemented in the Stata package. The analyses presented here are meant to be an illustration of the software and should not be considered a complete analysis of the Rotterdam breast cancer data. In the first example, we consider all death without distinguishing between death that happen before or after recurrence of the cancer. The entry into the intermediate state of recurrence will likely change the mortality rate, as well as introducing an additional time scale \textit{time since recurrence}. \citet{Batyrbekova:2022} includes a time-varying covariate to account for the entry into the intermediate state, and only introduce the second time scale after recurrence. The model implemented in \texttt{TwoTimeScales} is currently limited to the inclusion of fixed covariates only. 

The R-package \texttt{Epi} implements functions to work with data on the Lexis diagram, through the object \texttt{Lexis} and methods available for this type of object \citep{Plummer:2011,Epi} and is tailored to the analysis of epidemiological data. In \texttt{Epi} the time scales are also split in small intervals, and the log-hazard is modelled as a sum of baseline hazards over each of the time scales represented through natural splines. While the general setting of \texttt{Epi} is closely related to that of \texttt{TwoTimeScales}, the approach implemented in \texttt{Epi} does not include interactions between the time scales (see \citet{Iacobelli:2013} for more details), while such interactions being the main contribution of the model implemented by \texttt{TwoTimeScales}.

As future developments of the \texttt{TwoTimeScales} package analyses with time-dependent covariates and interval censored event times on multiple time scales are planned.

\section*{Conclusions}
In this paper we presented the R-package \texttt{TwoTimeScales} to analyse time-to-event data over two time scales simultaneously. We described the most important functions of the package and demonstrated their use to analyse a freely available dataset with follow-up data of patients with breast cancer. We showed that \texttt{TwoTimeScales} can be used to fit flexible hazard models with two time scales, producing new insights in the analysis of time-to-event data with multiple time scales. 

\bibliographystyle{apalike}

\bibliography{refs}

\newpage
\section{Appendix}
\label{appch5:A}

\subsection*{Further details of the fitting function}

\begin{figure}[!h]
   \begin{subfigure}[t]{0.48\textwidth}
        \centering
        \includegraphics[width=\textwidth]{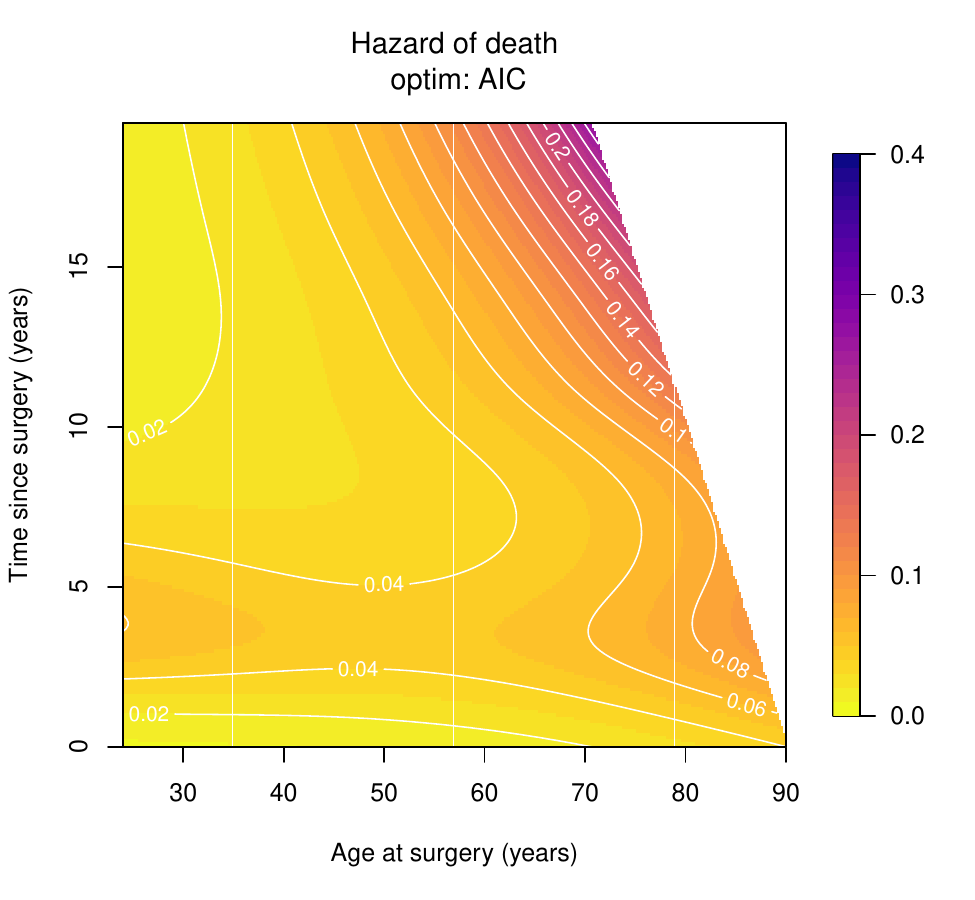}
   \end{subfigure}%
    ~~~ 
    \begin{subfigure}[t]{0.48\textwidth}
        \centering
        \includegraphics[width=\textwidth]{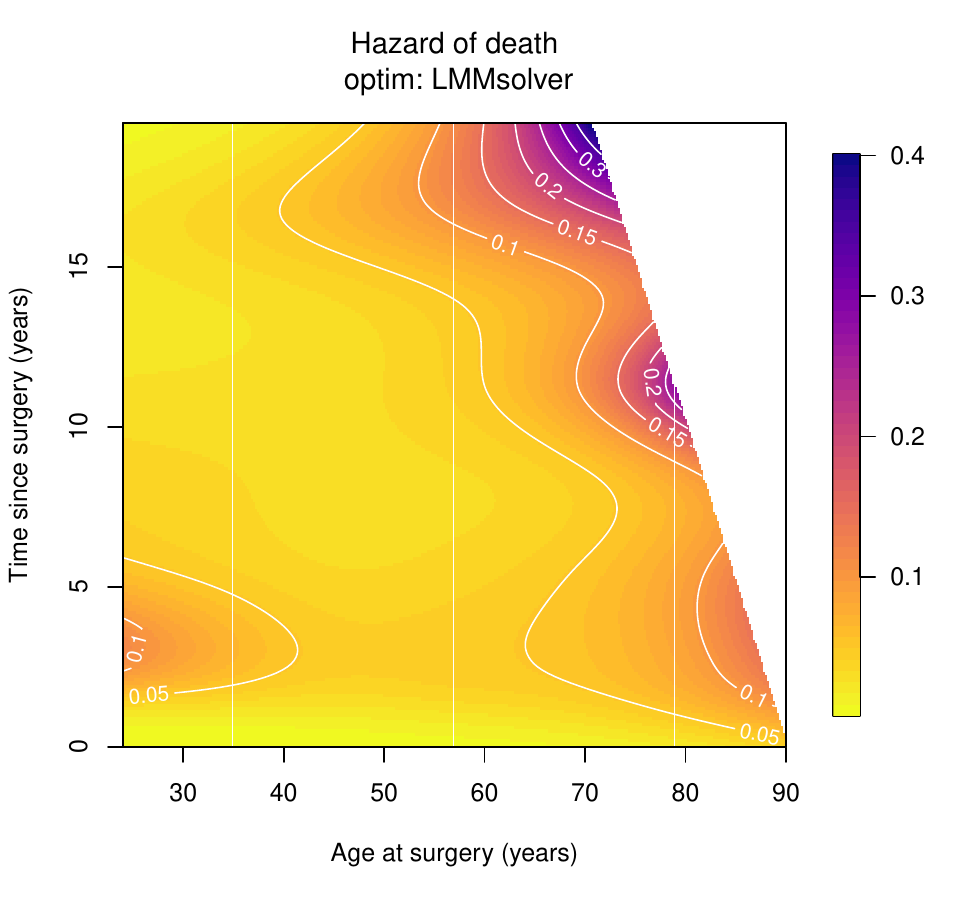}
    \end{subfigure}%

\caption{Left: Hazard of all deaths, smoothing parameters selected optimizing the AIC. Right: Hazard of all deaths, smoothing parameters obtained through mixed model representation.}
    \label{app:othermodels}
\end{figure}

In Section \textbf{Implementation} we have presented the most important arguments of the function \texttt{fit2ts()}, and showcased its use to fit a model for the hazard of dying after surgery, irrespective of recurrence status.
Here we will discuss some additional arguments of \texttt{fit2ts()}.
The user can provide information on the $B$-splines specification for the baseline hazard in a list \texttt{Bbases\_spec} with arguments \texttt{bdeg}, the degree of the $B$-splines bases, \texttt{nseg\_u} and \texttt{nseg\_s}, the number of segments for the B-splines over $u$ and $s$ respectively, \texttt{min\_u}, \texttt{max\_u}, \texttt{min\_s} and \texttt{max\_s} to specify the lower and upper limits of the $B$-splines bases over $u$ and $s$ respectively. The order of the penalty, which by default is two, can be specified in the argument \texttt{pord}. Three additional arguments control how the model is fitted and the optimal smoothing parameters are obtained. First, \texttt{optim\_method} selects whether a grid search over the smoothing parameters should be performed, numerical optimization should be used or the model should be estimated as a sparse linear mixed model through \texttt{LMMsolver}. The link to mixed model, as well as the use of the R-package \texttt{LMMsolver} is discussed in \cite{Carollo:2025}.  Second, unless the mixed model approach is used, the criterion that should be minimized (AIC or BIC) can be passed to the argument \texttt{optim\_criterion}. Lastly, the starting values for the smoothing parameters can be provided in the argument \texttt{lrho}. If \texttt{optim\_method = "grid\_search"}, the user can specify the values of $\log_{10}(\varrho_u)$ and $\log_{10}(\varrho_s)$ as elements in a list provided to the argument \texttt{lrho}. If the numerical optimization method is selected, the starting values can be provided to \texttt{lrho} as a vector with two elements. No starting values are needed if the model is estimated through \texttt{LMMsolver}. For the example presented in the paper, using \texttt{optim\_method = "LMMsolver"} results in a substantial gain in terms of computation time (2.46s), but overfits the data. We have opted for the model estimated via numerical optimization of the BIC, which also produces the most similar results to those presented in \citet{Bower:2022}. The model fit via the link to mixed method, as well as that obtained by optimizing the AIC are presented in the Figure~\ref{app:othermodels}.

\subsection*{Example 1: all deaths - other optimal models}

\end{document}